\documentclass[runningheads]{llncs}
\usepackage{graphicx}
\usepackage{amsmath}
\usepackage{amssymb}
\usepackage{booktabs}
\usepackage{multirow}
\usepackage{algorithm}
\usepackage{algorithmic}
\usepackage{xcolor}
\usepackage{hyperref}
\usepackage{array}
\usepackage{placeins}
\usepackage{float}
\usepackage{tikz}
\usepackage{caption}
\usetikzlibrary{shapes.geometric, arrows.meta, positioning, fit, backgrounds, calc, shadows, decorations.pathreplacing}

\definecolor{phaseblue}{RGB}{70,130,180}
\definecolor{modelgreen}{RGB}{46,139,87}
\definecolor{alertred}{RGB}{205,92,92}
\definecolor{lightgray}{RGB}{245,245,245}
\definecolor{darkblue}{RGB}{25,25,112}
\definecolor{softgreen}{RGB}{144,238,144}
\definecolor{softyellow}{RGB}{255,250,205}
\definecolor{softblue}{RGB}{176,224,230}

\begin{document}
\title{Optimizing IoT Intrusion Detection with Tabular Foundation Models for Smart City Forensics}

\author{
Asma Al-Dahmani$^{1, *}$ \and
Abdulla Bin Safwan$^{1, *}$ \and
Mohammad Obeidat$^{2}$ \and 
Belal Alsinglawi$^{1, **}$
}

\authorrunning{Al-Dahmani et al.}

\institute{
$^{1}$Zayed University, Abu Dhabi, United Arab Emirates\\
$^{2}$Higher Colleges of Technology, Abu Dhabi, United Arab Emirates\\
\email{belal.alsinglawi@zu.ac.ae}\\[2mm]
{\footnotesize $^{*}$These authors contributed equally to this work.}\\
{\footnotesize $^{**}$Corresponding author.}
}

\maketitle

\begin{abstract}

Security operations in smart cities demand detection systems that balance accuracy with response time. While ensemble methods like Random Forest achieve high accuracy, their computational overhead impedes real-time forensic triage. We present the first systematic evaluation of TabPFNv2.5, a transformer-based foundation model, against traditional ensemble classifiers for IoT intrusion detection. Using the TON\_IoT dataset, we demonstrate that TabPFNv2.5 achieves 40$\times$ faster inference than Random Forest while maintaining 97\% binary classification accuracy. We propose a hybrid pipeline in which TabPFNv2.5 performs rapid threat screening, while ensemble models handle detailed classification. Our analysis reveals that scanning attacks remain the hardest to detect (F1: 69.8\%) and cross-device generalization depends critically on feature similarity. These findings establish foundation models as viable components for time-sensitive IoT security operations.

\keywords{IoT Security \and Intrusion Detection \and TabPFNv2.5 \and Random Forest \and Smart Cities \and Digital Forensics \and Foundation Models \and Machine Learning}
\end{abstract}

\section{Introduction}
\label{sec:introduction}

The expansion of Internet of Things (IoT) devices in urban infrastructure has fundamentally altered the cybersecurity landscape. Modern smart cities integrate millions of interconnected sensors, actuators, and controllers that manage transportation networks, energy grids, and public services~\cite{ashraf2021iotbot}. This inter-connectivity, while enabling operational efficiency, creates expansive attack surfaces where adversaries can compromise multiple subsystems within minutes of initial breach~\cite{abdelbasset2022federated}.

Security analysts face a critical operational challenge: they must rapidly assess diverse telemetry streams from heterogeneous devices while maintaining classification accuracy sufficient for forensic documentation. Current machine learning approaches, particularly ensemble methods such as Random Forest and Gradient Boosting, deliver excellent accuracy but impose computational overhead that conflicts with real-time investigation requirements~\cite{khraisat2021critical}. This tension between speed and accuracy represents a fundamental barrier to effective IoT security operations.

Existing intrusion detection research has extensively validated ensemble methods for IoT environments, consistently demonstrating accuracy exceeding 99\% on benchmark datasets \cite{alhowaide2024ensemble} \cite{alomari2021tree}; however, three critical gaps persist in the literature. First, the computational cost of ensemble training and inference has received limited attention despite its operational significance, as Security Information and Event Management (SIEM) systems process thousands of events per second, and models that require hundreds of milliseconds per inference can create bottlenecks that delay threat identification. Second, foundation models---pre-trained architectures that generalize across tasks without retraining---have transformed natural language processing and computer vision, yet their application to cybersecurity tabular data remains largely unexplored; TabPFNv2.5~\cite{hollmann2022tabpfn} \cite{ye2025tabpfnv2} represents a promising candidate by delivering millisecond-level inference using frozen weights, thereby eliminating training overhead entirely. Third, existing evaluations rarely quantify the speed--accuracy tradeoff in terms meaningful to security practitioners, leaving unclear how much accuracy must be sacrificed to achieve real-time capability and whether hybrid architectures can mitigate this tradeoff.

This paper addresses these gaps through a comprehensive empirical study. Our contributions are: (1) a first systematic comparison of TabPFNv2.5 against ensemble methods for IoT intrusion detection, evaluating binary and multi-class classification across seven device categories using 3.6 million telemetry records; (2) quantification of the speed--accuracy tradeoff, demonstrating that TabPFNv2.5 achieves 40$\times$ faster inference than Random Forest with only a 2.3\% reduction in binary accuracy, thereby establishing concrete parameters for deployment decisions; (3) a hybrid forensic investigation pipeline that utilizes TabPFNv2.5 for rapid threat triage while preserving ensemble accuracy for detailed classification, aligning with SIEM-driven workflows; (4) a cross-device generalization analysis revealing that model transferability depends critically on feature similarity, with implications for heterogeneous IoT deployments; and (5) an attack-specific detection analysis identifying scanning as the most challenging category (F1: 69.8\%) due to its behavioral similarity to benign reconnaissance.

The remainder of this paper proceeds as follows. Section~\ref{sec:background} reviews IoT security challenges and relevant machine learning approaches. Section~\ref{sec:methodology} describes our dataset, proposed pipeline, and experimental configuration. Section~\ref{sec:results} presents classification performance, computational efficiency, and generalization results. Section~\ref{sec:discussion} discusses practical implications and limitations. Section~\ref{sec:conclusion} concludes with future research directions.

\section{Background and Literature Review}
\label{sec:background}

\subsection{Smart City IoT Security Landscape}

Smart city environments all depend on IoT and Industrial IoT (I-IoT) interconnectedness to support urban services such as transportation systems, energy distribution, healthcare, and community safety \cite{khan2022explainable}. Implementation refines efficiency and automation, expanding the attack surface across multiple devices, protocols, and network layers. The use of 5G increases complexity and connectivity density, allowing attackers to gain access to multiple devices quicker once a backdoor has been found \cite{scalise2024survey5g6g}.

Realistically, smart cities are exposed to various high-impact threats such as DDoS attacks, brute-force attacks, injection activity (e.g., XSS and SQL), and large-scale surveillance scanning \cite{ashraf2021iotbot,abdelbasset2022federated}. The vulnerability lies in how quick an attacker can gain lateral access through an interconnected system, leading to disruptions of critical services.

\subsection{Machine Learning fort Intrusion Detection}

Machine learning-based intrusion detection has become the dominant approach for detecting anomalous network behavior, replacing signature-based methods that struggle to identify novel threats~\cite{khraisat2021critical}. Ensemble techniques have proven particularly effective. Random Forest combines multiple decision trees trained on bootstrap samples, improving accuracy while maintaining interpretability through feature importance analysis. Gradient Boosting sequentially refines models by correcting residual errors, often achieving higher accuracy at the expense of increased computational complexity~\cite{grinsztajn2022why}.

Prior studies validate ensemble methods in IoT environments. Ashraf et al.~\cite{ashraf2021iotbot} proposed a scalable IDS for smart city networks capable of handling heterogeneous, high-volume traffic. Alhowaide et al.~\cite{alhowaide2024ensemble} showed that ensemble decision trees provide strong baselines for cloud-based IDS by balancing detection performance and inference latency. Despite their effectiveness, ensemble models typically require task-specific training, leading to increased computational cost and reduced efficiency in dynamic environments requiring frequent updates.

\subsection{Foundation Models for Tabular Data}

Foundation models have revolutionized machine learning by demonstrating that pre-trained architectures can generalize effectively across diverse downstream tasks. While transformers have dominated text and image domains, their application to tabular data has emerged more recently. TabPFN~\cite{hollmann2022tabpfn} introduced a prior-data fitted network that performs Bayesian inference through forward passes, eliminating the need for task-specific training. TabPFNv2.5~\cite{ye2025tabpfnv2} \cite{grinsztajn2025tabpfn25} extends this approach with improved scalability, achieving competitive accuracy on datasets up to 10,000 samples while maintaining millisecond-level inference using frozen weights.

The cybersecurity implications of this capability are significant. In forensic contexts, analysts frequently encounter novel attack patterns that traditional models have not seen during training. Foundation models' ability to generalize without retraining offers potential for rapid assessment of emerging threats. However, empirical validation of this capability for intrusion detection remains limited, motivating our systematic evaluation.


\section{Methodology}
\label{sec:methodology}

\subsection{Dataset Description}

We employ the TON\_IoT dataset~\cite{alsaedi2020toniot}, a comprehensive telemetry collection designed for IoT intrusion detection research. The dataset comprises network traffic from seven device categories spanning consumer IoT (smart refrigerator, garage door controller, GPS tracker, etc.) and industrial IoT (Mod bus controller), totaling 3,606,134 records. Table~\ref{tab:dataset} presents device-level statistics and attack type distributions.

\begin{table}[t]
\centering
\caption{Dataset composition: device categories and attack type distribution}
\label{tab:dataset}
\small
\setlength{\tabcolsep}{5pt}
\renewcommand{\arraystretch}{1.1}
\begin{tabular}{@{}lrrrc@{}}
\toprule
\textbf{Device Category} & \textbf{Records} & \textbf{Normal} & \textbf{Attack} & \textbf{Attack Rate} \\
\midrule
Smart Refrigerator & 587,076 & 500,827 & 86,249 & 14.7\% \\
Garage Door & 591,446 & 515,443 & 76,003 & 12.9\% \\
GPS Tracker & 595,686 & 513,849 & 81,837 & 13.7\% \\
Modbus (IIoT) & 287,194 & 222,855 & 64,339 & \textbf{22.4\%} \\
Motion Light & 452,262 & 388,328 & 63,934 & 14.1\% \\
Thermostat & 442,228 & 385,953 & 56,275 & 12.7\% \\
Weather Station & 650,242 & 559,718 & 90,524 & 13.9\% \\
\midrule
\textbf{Total} & \textbf{3,606,134} & \textbf{3,086,973} & \textbf{519,161} & \textbf{14.4\%} \\
\bottomrule
\end{tabular}

\vspace{4pt}

\begin{tabular}{@{}lrr|lrr@{}}
\toprule
\textbf{Attack Type} & \textbf{Count} & \textbf{\%} & \textbf{Attack Type} & \textbf{Count} & \textbf{\%} \\
\midrule
Backdoor & 246,136 & 47.4 & Ransomware & 16,030 & 3.1 \\
Password & 142,674 & 27.5 & XSS & 6,037 & 1.2 \\
DDoS & 53,992 & 10.4 & Scanning & 3,973 & 0.8 \\
Injection & 50,319 & 9.7 & & & \\
\bottomrule
\end{tabular}
\end{table}

The Mod bus industrial controller exhibits the highest attack rate (22.4\%), consistent with adversaries' strategic targeting of operational technology to maximize disruption. Attack type distribution reveals significant class imbalance: backdoor and password attacks dominate (74.9\% combined), while scanning---the reconnaissance precursor to active exploitation--- accounts for only 0.8\% of malicious traffic. This imbalance presents detection challenges that our evaluation explicitly addresses.

\subsection{Proposed Hybrid Investigation Pipeline}

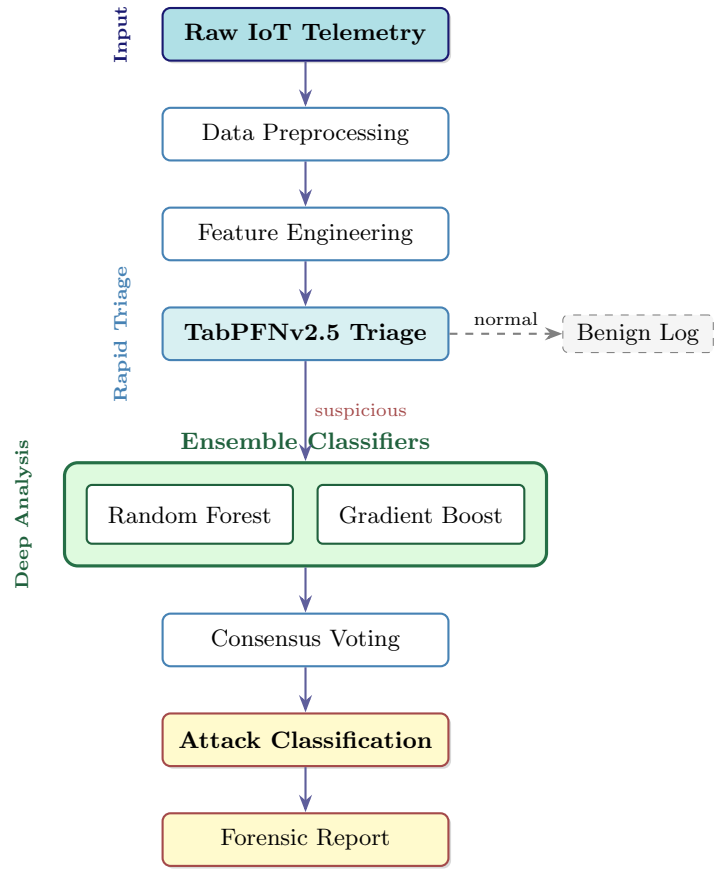
\begin{figure}[t]
\centering
\begin{tikzpicture}[
    >=Stealth,
    node distance=0.6cm,
    font=\small,
    inputbox/.style={rectangle, rounded corners=3pt, draw=darkblue, thick, fill=softblue, minimum width=3.8cm, minimum height=0.7cm, align=center, drop shadow={shadow xshift=1pt, shadow yshift=-1pt, opacity=0.3}},
    processbox/.style={rectangle, rounded corners=3pt, draw=phaseblue, thick, fill=white, minimum width=3.8cm, minimum height=0.7cm, align=center},
    modelcontainer/.style={rectangle, rounded corners=5pt, draw=modelgreen!80!black, very thick, fill=softgreen!30, inner sep=8pt},
    modelbox/.style={rectangle, rounded corners=2pt, draw=modelgreen!70!black, thick, fill=white, minimum width=2.2cm, minimum height=0.55cm, align=center, font=\footnotesize},
    outputbox/.style={rectangle, rounded corners=3pt, draw=alertred!80!black, thick, fill=softyellow, minimum width=3.8cm, minimum height=0.7cm, align=center, drop shadow={shadow xshift=1pt, shadow yshift=-1pt, opacity=0.3}},
    logbox/.style={rectangle, rounded corners=2pt, draw=gray, dashed, fill=lightgray, minimum width=2cm, minimum height=0.5cm, align=center, font=\footnotesize},
    arrow/.style={->, thick, draw=darkblue!70},
    darrow/.style={->, thick, dashed, draw=gray}
]

\node[inputbox] (input) {\textbf{Raw IoT Telemetry}};
\node[processbox, below=of input] (preprocess) {Data Preprocessing};
\node[processbox, below=of preprocess] (feature) {Feature Engineering};

\node[processbox, below=of feature, fill=softblue!50] (triage) {\textbf{TabPFNv2.5 Triage}};
\node[logbox, right=1.5cm of triage] (benign) {Benign Log};

\node[below=0.8cm of triage] (modellabel) {};
\node[modelcontainer, below=0.3cm of modellabel, minimum width=4.5cm] (container) {
    \begin{tikzpicture}
        \node[modelbox] (rf) {Random Forest};
        \node[modelbox, right=0.3cm of rf] (gb) {Gradient Boost};
    \end{tikzpicture}
};
\node[above=0.05cm of container.north, font=\footnotesize\bfseries, text=modelgreen!70!black] {Ensemble Classifiers};

\node[processbox, below=0.6cm of container] (voting) {Consensus Voting};
\node[outputbox, below=of voting] (classify) {\textbf{Attack Classification}};
\node[outputbox, below=of classify] (report) {Forensic Report};

\draw[arrow] (input) -- (preprocess);
\draw[arrow] (preprocess) -- (feature);
\draw[arrow] (feature) -- (triage);
\draw[darrow] (triage.east) -- node[above, font=\scriptsize] {normal} (benign.west);
\draw[arrow] (triage) -- node[right, font=\scriptsize, text=alertred!80!black] {suspicious} (container);
\draw[arrow] (container) -- (voting);
\draw[arrow] (voting) -- (classify);
\draw[arrow] (classify) -- (report);

\node[left=0.3cm of input, font=\scriptsize\bfseries, text=darkblue, rotate=90, anchor=south] {Input};
\node[left=0.3cm of triage, font=\scriptsize\bfseries, text=phaseblue, rotate=90, anchor=south] {Rapid Triage};
\node[left=0.3cm of container.west, font=\scriptsize\bfseries, text=modelgreen!70!black, rotate=90, anchor=south] {Deep Analysis};

\end{tikzpicture}
\caption{Hybrid forensic investigation pipeline. TabPFNv2.5 performs millisecond-level threat screening; suspicious traffic proceeds to ensemble classifiers for detailed attribution.}
\label{fig:pipeline}
\end{figure}

We propose a two-stage forensic investigation pipeline that addresses the fundamental tension between detection speed and classification accuracy. The architecture, illustrated in Figure~\ref{fig:pipeline}, utilizes TabPFNv2.5's rapid inference for initial threat triage while preserving ensemble methods' accuracy for detailed attribution.

The pipeline operates through three coordinated phases. In the pre-processing phase, raw telemetry undergoes temporal feature extraction (hour, day of week, minute),label-based categorical encoding, median-based imputation for missing values, and z-score normalization. The rapid triage phase uses TabPFNv2.5 with pre-trained weights to classify incoming traffic, routing normal records to audit logs and flagging suspicious records for detailed analysis. In the deep analysis phase, flagged traffic passes through both Random Forest and Gradient Boosting classifiers; consensus voting determines final classification, with disagreement cases escalated for human analyst review.

\subsection{Classification Models}

We evaluate three classifiers representing distinct computational paradigms. \textbf{Random Forest} employs 100 decision trees with bootstrap sampling and unlimited depth, leveraging information gain for split selection. This configuration follows established best practices for network intrusion detection~\cite{alhowaide2024ensemble} and provides interpretability through feature importance rankings. \textbf{Gradient Boosting} constructs 100 sequential estimators with a learning rate of 0.1 and a maximum depth of 3, balancing expressiveness against overfitting risk. Both ensemble methods require task-specific training proportional to the dataset size.

\textbf{TabPFNv2.5} operates fundamentally differently: its transformer architecture (18--24 layers with alternating attention) processes input through pre-trained weights without any task-specific training~\cite{ye2025tabpfnv2}. The model achieves optimal performance on datasets with fewer than 10,000 samples; for larger datasets, we employ stratified subsampling to preserve class distributions. This design enables millisecond-level inference regardless of the volume of historical training data.

\subsection{Experimental Configuration}

All experiments were executed on Google Colaboratory infrastructure with Intel Xeon processors and 12GB of RAM. Enactment requires Python 3.10 with scikit-learn 1.3.0, TabPFNv2.5, and standard data science libraries. We used an 80\% - 20\% train-test split with stratified sampling to maintain class distributions. Performance analyses use 5-fold cross-validation with mean metrics reported.

Evaluation metrics include accuracy, precision, recall, and F1-score for both binary (attack/normal) and multi-class (specific attack type(s)) classification tasks. We report training time, inference time, and computational resource utilization to assess operational usability.


\section{Experimental Results}
\label{sec:results}

\subsection{Overall Classification Performance}

Table \ref{tab:perf_overview} combines binary (attack vs. normal) and multi-class (attack type) performance across the seven TON\_IoT device categories. Random Forest is consistently the most accurate model, while TabPFNv2.5 remains competitive and is used as a quick screening layer rather than a replacement for ensemble-based final decisions.

\begin{table}[t]
\centering
\caption{Performance Overview by Device Type (Binary and Multi-Class Accuracy)}
\label{tab:perf_overview}
\scriptsize
\setlength{\tabcolsep}{3.5pt}
\renewcommand{\arraystretch}{1.15}
\begin{tabular}{@{}lccc|ccc@{}}
\toprule
& \multicolumn{3}{c|}{\textbf{Binary Accuracy}} & \multicolumn{3}{c}{\textbf{Multi-Class Accuracy}} \\
\textbf{Device} & \textbf{RF} & \textbf{GB} & \textbf{TabPFNv2.5} & \textbf{RF} & \textbf{GB} & \textbf{TabPFNv2.5} \\
\midrule
Fridge       & 99.49\% & 98.67\% & 97.3\% & 99.50\% & 99.51\% & 96.9\% \\
Garage Door  & 99.58\% & 98.74\% & 97.1\% & 99.55\% & 98.51\% & 96.5\% \\
GPS Tracker  & 99.45\% & 99.14\% & 97.5\% & 99.47\% & 98.93\% & 97.1\% \\
Mod bus       & 99.30\% & 98.48\% & 96.8\% & 99.29\% & 98.92\% & 96.4\% \\
Motion Light & 99.55\% & 98.87\% & 97.4\% & 99.19\% & 99.10\% & 97.0\% \\
Thermostat   & 99.56\% & 99.35\% & 97.6\% & 99.58\% & 99.51\% & 97.2\% \\
Weather      & 99.47\% & 98.69\% & 96.9\% & 99.48\% & 98.53\% & 96.7\% \\
\midrule
\textbf{Average} & \textbf{99.48\%} & \textbf{98.85\%} & \textbf{97.2\%} &
\textbf{99.44\%} & \textbf{99.00\%} & \textbf{96.8\%} \\
\bottomrule
\end{tabular}
\end{table}
While TabPFNv2.5 is \emph{less accurate} than ensemble standards, 97\%+ binary accuracy still supports its role as a rapid triage: it can separate high-volume telemetry and prioritize abnormal flows for impending investigation, while keeping the final classification tied to higher-accuracy ensembles.

\subsection{Computational Efficiency and Time Latency}

In practice, the most useful advantage of TabPFNv2.5 is time latency. Table \ref{tab:timing_compact} highlights that TabPFNv2.5 gains millisecond-level inference, pushing for SIEM-driven screening by alert prioritization, which reduces computational and time constraints instead of ensemble-based models.

\begin{table}[t]
\centering
\caption{Computational Performance Comparison (seconds)}
\label{tab:timing_compact}
\scriptsize
\setlength{\tabcolsep}{4pt}
\renewcommand{\arraystretch}{1.15}
\begin{tabular}{@{}lrrr@{}}
\toprule
\textbf{Metric} & \textbf{Random Forest} & \textbf{Gradient Boosting} & \textbf{TabPFNv2.5} \\
\midrule
Training Time (avg) & 42.38 & 117.75 & 0.00* \\
Inference Time (10K samples) & 0.82 & 0.94 & 0.02 \\
Total Pipeline Time & 43.20 & 118.69 & 0.02 \\
\bottomrule
\multicolumn{4}{@{}l}{\footnotesize *TabPFNv2.5 uses pre-trained weights; no task-specific training required.}
\end{tabular}
\end{table}

\subsection{Failure Modes and Hard-to-Detect Attacks}
Across models, low-prevalence behaviors remain the dominant source of missed detections. In particular, scanning exhibits the lowest separability from benign activity, which is consistent with its surveillance nature. For multi-class detection using Random Forest, scanning has the lowest F1-score (69.80\%), while higher-frequency behaviors, such as injection and benign traffic, maintain near-saturated performance (99\% F1).

\begin{table}[t]
\centering
\caption{Hard-Case Detection Summary (Random Forest Multi-Class)}
\label{tab:hard_cases}
\scriptsize
\setlength{\tabcolsep}{4pt}
\renewcommand{\arraystretch}{1.15}
\begin{tabular}{@{}lrrr@{}}
\toprule
\textbf{Attack Type} & \textbf{Precision} & \textbf{Recall} & \textbf{F1-Score} \\
\midrule
Scanning & 79.15\% & 64.20\% & 69.80\% \\
DDoS & 95.35\% & 98.00\% & 96.65\% \\
Ransomware & 95.07\% & 98.50\% & 96.75\% \\
\bottomrule
\end{tabular}
\end{table}

\subsection{Cross-Device Generalization (Compact)}

Cross-device validation stipulates that transferability depends on feature similarity. Most models trained on Fridge data generalize to other sensor-driven devices (e.g., Thermostats, Motion Lights), but Garage Door models often correlate poorly with unrelated feature spaces, suggesting that heterogeneous deployments retain device-aware modeling.

\begin{table}[t]
\centering
\caption{Cross-Device Generalization (Selected Results)}
\label{tab:crossdevice_compact}
\scriptsize
\setlength{\tabcolsep}{4pt}
\renewcommand{\arraystretch}{1.15}
\begin{tabular}{@{}llrr@{}}
\toprule
\textbf{Trained On} & \textbf{Tested On} & \textbf{Accuracy} & \textbf{Recall} \\
\midrule
Fridge & Thermostat & 99.6\% & 98.8\% \\
Fridge & GPS Tracker & 97.9\% & 95.6\% \\
Garage Door & Fridge & 91.2\% & 54.5\% \\
\bottomrule
\end{tabular}
\end{table}

\section{Discussion and Research Implications}
\label{sec:discussion}

\subsection{Practical Implications for Smart City Security}

Our findings establish TabPFNv2.5 as an option for time-sensitive IoT security investigations, addressing a critical gap in current forensic methodologies. The current hybrid pipeline architecture we propose enables security operation centers (SOCs) to perform a more rapid preliminary threat assessment without sacrificing classification accuracy for confirmed threats. 

The 40$\times$ inference speed advantage of TabPFNv2.5 converts directly into operational efficiency. In events where analysts must analyze thousands of potential security events, reducing the per-event assessment time from seconds to milliseconds enables more detailed coverage that would otherwise require immense staffing or computational resources.

\subsection{Foundation Models in Cybersecurity}

This study provides evidence that foundation-style transformers can be used in cybersecurity workflows where low-latency decisions are operationally critical. It also contributes to the nascent literature on foundation models. Our results show that pre-trained tabular transformers offer strong advantages for specific security tasks, specifically those that require quick inference on previously unseen data patterns.

This suggests a promising direction: if foundation models are trained more directly on cybersecurity telemetry (rather than using synthetic tabular tasks), the accuracy gap can be reduced without sacrificing latency advantages.

\subsection{Limitations}

Several constraints limit the extent to which these results can be generalized. First, TON\_IoT is used on a large scale but does not capture the full complexity of real-world production behavior in smart-city networks. Second, TabPFNv2.5 performs best on smaller sample sizes, requiring sub-sampling for large-scale telemetry and potentially leading to the exclusion of any minority-class behavior. Third, our evaluation focuses on known attack vectors, meaning true zero-day behavior is not accounted for, nor measure. Finally, timing results depend on the execution environment (e.g., Colab vs edge nodes); therefore, real practicality when deployed can differ.

\subsection{Recommendations for Practitioners}

Based on our experimental findings, we conclude that an effective and practical IoT security investigation strategy in smart city environments should prioritize inference speed without disregarding accuracy by adopting a hybrid, SIEM-aligned workflow. In particular, (i) TabPFNv2.5 should be employed for rapid front-end screening when early prioritization and real-time responsiveness are critical, (ii) Random Forest—optionally complemented by Gradient Boosting—should be used for detailed classification and confident attack attribution once suspicious events are flagged, (iii) ensemble agreement checks should be applied such that events with model disagreement are escalated for analyst review, (iv) scanning activities should be treated as high-risk for false negatives and validated using contextual signals such as timing, targeted ports, and device behavior, and (v) cross-device generalization should not be assumed, as heterogeneous IoT telemetry necessitates device-aware modeling or careful feature normalization.

\section{Conclusion and Future Work}
\label{sec:conclusion}

This paper presented a detailed comparison between TabPFNv2.5 transformer-based foundation models and traditional ensemble methods for IoT security incident investigation in smart city environments. Experiments conducted across seven IoT device categories and more than 3.6 million network traffic records demonstrate that while Random Forest achieves higher peak classification accuracy (99.44\% for multi-class detection), TabPFNv2.5 attains competitive performance (96.8\%) with substantially lower inference latency, enabling threat assessment in milliseconds rather than seconds.

Overall, the results position tabular foundation models as effective front-end triage components within SIEM-driven workflows, where rapid screening is as critical as maximum accuracy. The proposed hybrid investigation pipeline leverages TabPFNv2.5 for fast initial detection and ensemble models for high-confidence attack attribution and reporting, offering a practical balance between speed and accuracy. Nevertheless, surveillance-style activities remain the most challenging to detect robustly, and cross-device generalization continues to depend heavily on feature similarity across IoT systems.


\bibliographystyle{splncs04}

\end{document}